\documentclass[%
 reprint,
showpacs,preprintnumbers,
 amsmath,amssymb,
 aps, showkeys
]{revtex4-1}
\usepackage{color}
\usepackage{graphicx}
\usepackage{dcolumn}
\usepackage{bm}
\usepackage{hyperref}

\newcommand\beq{\begin{equation}}
\newcommand\eeq{\end{equation}}
\newcommand\beqs{\begin{equation*}}
\newcommand\eeqs{\end{equation*}}
\newcommand\beqa{\begin{eqnarray}}
\newcommand\eeqa{\end{eqnarray}}
\newcommand\beqas{\begin{eqnarray*}}
\newcommand\eeqas{\end{eqnarray*}}

\begin{document}

\title{A Reissner-Nordstr\"{o}m$+\Lambda$ black hole in the
Friedman-Robertson-Walker universe}

\author{Safiqul Islam}
\email{sofiqul001@yahoo.co.in}
\affiliation{Department of Mathematics, St. Theresa International College, Thailand}

\author{Priti Mishra}
\email{preet.tifr@gmail.com}
\affiliation{Department of Physics, Magadh Mahila College, Patna University, India}

\author{Somi Aktar}
\email{somiaktar9@gmail.com}
\affiliation{Department of Mathematics,Jadavpur University,Kolkata-700 032,West Bengal,India}

\author{Farook Rahaman}
\email{rahaman@associates.iucaa.in}
\affiliation{Department of Mathematics,Jadavpur University,Kolkata-700 032,West Bengal,India}

\date{\today}
\begin{abstract}
A charged, non-rotating, spherically symmetric black hole which has
cosmological constant $\Lambda$ (Reissner-Nordstr\"{o}m+$\Lambda$ or
RN+$\Lambda$),  active gravitational  mass $M$ and electric charge $Q$ is  studied in exterior Friedman-Robertson-Walker (FRW) universe in
(2+1) dimensional spacetime. We find new class of exact solutions of the charged black hole. It is found that the cosmological constant is negative inside the black hole. We
confirm it from the geodesic equations too. The cosmological constant is found to be dependent on
$R$, $Q$ and $a(v)$ which correspond to the areal radius, charge, of the black hole and the scale factor of
the universe respectively. We note that the expansion of the universe affects the size and the mass
of the black hole. An important observation is that, for an observer at infinity, both the mass and
charge of black hole increase with the contraction of the universe and decrease with the expansion
of the universe.
\end{abstract}


\maketitle

\section{Introduction:}\label{intro}
Ever since their advent black holes have been studied in a
great detail. However, almost all previous studies have focused either on
isolated or binary black holes. But in reality  black holes are neither
isolated nor only in binaries. They are actually embedded in the background of
expanding universe. Therefore, we must study black holes in non-flat
backgrounds in order to understand the black holes in real universe.

The main motivation to study ($2+1$)-dimensional spacetimes admitting black hole solutions is that the $d=3$ cases have now attracted more attention and interest as compared to other ($d\geq4$) spacetimes with special mass and charge dependence. The solution of the Einstein field equations in ($2+1$)-d exhibit many characteristics of the ($3+1$)-d black hole. Moreover the structure of ($2+1$)-d black hole is simple enough to derive a number of exact computations, particularly in the quantum realm and string theory, which are not possible in ($3+1$) dimensions. Such study may also give us a way to unify gravity and quantum theory. We know that the entropy of a black hole is proportional to its surface area, which is also consistent with a ($2+1$)-d black hole. We are further interested to derive the equation of motion for geodesics in vicinity of spacetime of a ($2+1$)-dimensional charged black hole.
		
It is well known also that at the core or center of a black hole, according to general relativity, is a gravitational singularity, which is indeed a one-dimensional point. Its huge mass is located in an infinitesimal small space, where density as well as gravity is infinite and space-time curves infinitely. Hence, the notion of an actual physical singularity appears quite unlikely, and possibly points to general relativity being rather incomplete.

The work on ($2+1$)-d gravity theories has seen a great increase after the discovery that ($2+1$)-d general relativity possesses a black hole solution [Banados et al., 1992 \cite{f}]. It is the first example in this regard. The authors have observed that the fascinating properties of classical and especially quantum black hole, have long made it desirable to work on a lower-dimensional analog which could exhibit the key features and avert the unnecessary complications. It has been observed that such analog does exist in standard ($2+1$)-d Einstein-Maxwell theory with a negative cosmological constant. This has further motivated us to work on ($2+1$)-d black hole including the cosmological constant inside it.

Later on, Einstein-Maxwell \cite{i} and Einstein-Maxwell-dilaton \cite{j} extensions were also found. The authors studied black hole solutions which include all special characteristics that are observed in ($3+1$) or higher dimensional black holes like horizon(s), black hole thermodynamics as well as Hawking radiation.

The dimensional reduction of black hole solutions in 4D general relativity is done and new 3D black hole solutions with an isotropic event horizon are obtained by Zanchin et al. \cite{g}. Such event horizon is a typical characteristic of black hole and is an important study in our research too. The authors considered a 4D spacetime with one spacelike Killing vector and observed that it is possible to split the Einstein-Hilbert-Maxwell action with a cosmological term in terms of 3D quantities. The authors in \cite{h} have further formulated the three-dimensional Einstein-Maxwell-dilaton theory from the usual four-dimensional Einstein-Maxwell-Hilbert action for general relativity and observed that the 3D static spherically symmetric solution is analogous to the 4D Reissner-Nordstr$\ddot{o}$m-AdS black hole.
A particular case of the 3D action which presents Maxwell field, dilaton field and an extra scalar field, besides gravity field and a negative cosmological constant is chosen by them, and new 3D static black hole solutions are obtained.

In \cite{k}, the authors have constructed a large class of black hole solutions   by the power Maxwell field. Here the Maxwell scalar has the form $(F_{\mu \nu} F^{\mu \nu})^k$. The particular choice of $k=\frac{d}{4}$, yields in general a traceless Maxwell's energy-momentum tensor. They however observed that $k=\frac{3}{4}$ yields a general solution in ($2+1$)-dimensional Einstein-power-Maxwell (EPM) spacetime which is devoid of the traceless condition.

The references are much more, but limited here, which have aroused the interest to study the Reissner-Nordstr\"{o}m$+\Lambda$ black hole in the
Friedman-Robertson-Walker universe.

The study on black holes is not completely new. It started long back in 1933 when McVittie
\cite{mcvittie33} obtained his celebrated metric for a mass-particle in an
expanding universe. This metric is nothing but the Schwarzschild black hole
which is embedded in the Friedman-Robertson-Walker universe. In 1993, Kastor and
Traschen found the multi-black holes solution in the background of de Sitter
universe \cite{kastor93a,kastor93b}. The Kastor-Traschen solution describes the
dynamical system of arbitrary number of extreme Reissner-Nordstrom black holes
in the background of de Sitter universe. In 1999, Shiromizu and Gen studied
charged rotating black hole in de Sitter background
\cite{shiromizu2000}. In 2000, Nayak et al. studied the solutions for the
Schwarzschild and Kerr black holes in the
background of the Einstein universe \cite{nayak2000,nayak2000report}. In
2004 Gao et al. studied Reissner-Nordstr$\ddot{o}$m black hole in the
expanding universe \cite{gao04}.

In this paper, we extend the above studies from charged black holes into
charged black holes which have cosmological constant inside them. It has been
found in the literature that there are three possible black hole solutions
depending on the cosmological constant being (1) positive (2)
negative and (3) zero \cite{bousso12}.
We first deduce the metric for a
Reissner-Nordstr$\ddot{o}$m+$\Lambda$ black hole in the expanding universe. We
show that several special cases of our solution are exactly the same as some
solutions discovered previously.  We then study the effects
of the evolution of the universe on the size, mass and charge of the black hole.

We know that black holes exert a strong
gravitational influence due to their mass, just like every other massive object
in the Universe. This is how we actually discover and measure the mass of black
holes, by watching their effect through gravitational lensing, accretion, X-ray emissions etc.  For instance, the
supermassive black hole at the center of the Milky Way galaxy is so strong
gravitationally that the stars very near it orbit at a very, very high rate.
Using this and the equations that describe the orbits of these stars, we can
actually estimate the mass of the black hole.

N. Kaloper et al. \cite{kaloper10} has analyzed the McVittie solutions
of Einstein’s field equations for describing the gravitational fields of
spherically symmetric mass distributions in expanding FRW universes. They
focused on spatially flat McVittie geometries and showed that the McVittie
solutions which asymptote to FRW universes and dominated by a positive
cosmological constant at late times are black holes with regular event
horizons. Near the hole the charge contributions correct the effective
potential for the scalar and give it a large mass, as the supersymmetric
attractor mechanism in asymptotically flat black holes.

T. Maki et al. \cite{maki93} have studied (N + 1)-dimensional cosmological
solutions describing the multi-black hole configuration in the same system with
a cosmological constant. They investigated that the cosmological evolution of
the scale factor depends on the coupling of the dilaton to the cosmological
constant. The outline of our paper is envisaged as follows:

In section \ref{sec2} we have solved the Einstein-Maxwell field equations for the
static spherically symmetric line element for interior spacetime of a RN+$\Lambda$
black hole. The event horizons have been studied. The pressure, matter density and proper
charge density of the black hole has been expressed in terms of the mass, charge and the cosmological constant $(\Lambda)$.
The geodesics have been further verified. In section \ref{sec3} we transform the
RN+$\Lambda$ metric to the McVittie form \cite{mcvittie33} under suitable
transformation conditions for compatible study with respect to the
FRW universe. The various subconditions are specified. In section \ref{sec4} the boundary
conditions are discussed and we
further confirm the negative value of cosmological constant inside the black
hole. The value of the curvature parameter $k$ in the FRW metric is discussed. That the transformed RN+$\Lambda$ metric is an exact solution of the $EM$ field equations and the metric is physically relevant has been studied in section \ref{sec5} . We discuss the Darmois-Israel matching conditions in section \ref{sec6}. In section \ref{sec7} we further study the surface continuity. The study ends with a concluding remark in section \ref{conclusions} .

\section{Interior Reissner-Nordstr\"{o}m with $\Lambda$ metric:}\label{sec2}

We know that if an electrically charged particle falls into the Schwarzschild
black hole it
becomes charged. To describe such a black hole one has to solve the
Einstein-Maxwell equations considering the stress-energy tensor of the
electromagnetic field. RN+$\Lambda$  metric is a static
solution to the Einstein-Maxwell field equations, which corresponds to the
electrovacuum
gravitational field of a charged, non-rotating, spherically symmetric black hole
of mass M. Hence, we follow the analogue of the RN+$\Lambda$ solution
with exterior FRW metric for a spacetime with a cosmological constant.
Under such conditions, the metric of the line element for the interior space-time of
a static spherically symmetric charged distribution of matter in $(2+1)$ dimensions is of the form,

\begin{eqnarray}
ds^{2}&=&-(1-\frac{2M}{r}+\frac{Q^2}{r^2}-\frac{{\Lambda}r^2}{3})dt^2
\nonumber\\
      & &+(1-\frac{2M}{r}+\frac{Q^2}{r^2}-\frac{{\Lambda}r^2}{3})^{-1}dr^2
\nonumber\\
      & &+r^{2} d\theta^{2},
      \label{rnlmetric}
\end{eqnarray}

where M and Q are the mass and charge of the black hole, respectively and
$\Lambda$, the cosmological constant.

We have included the charge $Q$ inside the metric. Unlike our $2+1$-d metric, we find that the $3+1$-d metric in \cite{kottler}, is devoid of any charge.

{
The coordinate speed of light signal [Null geodesic] is obtained with $ds^2=0$ , hence we obtain
from eqn.(\ref{rnlmetric}),

\begin{eqnarray}
0&=&-(1-\frac{2M}{r}+\frac{Q^2}{r^2}-\frac{{\Lambda}r^2}{3})dt^2 \nonumber\\
      & &+(1-\frac{2M}{r}+\frac{Q^2}{r^2}-\frac{{\Lambda}r^2}{3})^{-1} dr^2
\nonumber\\
      & &+r^{2} d\theta^{2},
\end{eqnarray}

This implies,
\begin{eqnarray}
(\frac{dr}{dt}^2)&=&(1-\frac{2M}{r}+\frac{Q^2}{r^2}-\frac{{\Lambda}r^2}{3})
\nonumber\\
                 &
&.[(1-\frac{2M}{r}+\frac{Q^2}{r^2}-\frac{{\Lambda}r^2}{3})-r^{2}(\frac{d\theta}{
dt})^2],
\end{eqnarray}

At the surface $r=R$ on which $\frac{dr}{dt}=0$ (i.e on the RN+$\Lambda$ blackhole
surface), light cannot escape from this black hole surface,thus,
\begin{equation}
1-\frac{2M}{R}+\frac{Q^2}{R^2}-\frac{{\Lambda}R^2}{3}=0,
\end{equation}

Besides the cosmological constant the charged black hole is characterized by
two parameters, the mass M and the electric charge Q.} $\Lambda=0$ corresponds to
the
Reissner Nordstr\"{o}m metric \cite{frolovbook}, which is not our
case.
\subsection{Horizons in the RN+$\Lambda$ spacetime:}
On solving the above eqn.(4), { we get four values of the radial parameter, $r$,} given by,
\begin{eqnarray}
{r_1} &=&\frac{1}{2}.[\frac{2}{\Lambda}-\frac{a}{\Lambda
b}-\frac{b}{c}]^{\frac{1}{2}} \nonumber\\
      & &-\frac{1}{2}.[\frac{4}{\Lambda}+\frac{a}{\Lambda
b}+\frac{b}{c}-\frac{12 M}{\Lambda (\frac{2}{\Lambda}-\frac{a}{\Lambda
b}-\frac{b}{c})^\frac{1}{2}}]^\frac{1}{2},
\end{eqnarray}

\begin{eqnarray}
{r_2} &=&\frac{1}{2}.[\frac{2}{\Lambda}-\frac{a}{\Lambda
	b}-\frac{b}{c}]^{\frac{1}{2}} \nonumber\\
& &+\frac{1}{2}.[\frac{4}{\Lambda}+\frac{a}{\Lambda
	b}+\frac{b}{c}-\frac{12 M}{\Lambda (\frac{2}{\Lambda}-\frac{a}{\Lambda
		b}-\frac{b}{c})^\frac{1}{2}}]^\frac{1}{2},
\end{eqnarray}

\begin{eqnarray}
{r_3} &=&-\frac{1}{2}.[\frac{2}{\Lambda}-\frac{a}{\Lambda
	b}-\frac{b}{c}]^{\frac{1}{2}} \nonumber\\
& &-\frac{1}{2}.[\frac{4}{\Lambda}+\frac{a}{\Lambda
	b}+\frac{b}{c}-\frac{12 M}{\Lambda (\frac{2}{\Lambda}-\frac{a}{\Lambda
		b}-\frac{b}{c})^\frac{1}{2}}]^\frac{1}{2},
\end{eqnarray}
and
\begin{eqnarray}
{r_4} &=&-\frac{1}{2}.[\frac{2}{\Lambda}-\frac{a}{\Lambda
	b}-\frac{b}{c}]^{\frac{1}{2}} \nonumber\\
& &+\frac{1}{2}.[\frac{4}{\Lambda}+\frac{a}{\Lambda
	b}+\frac{b}{c}-\frac{12 M}{\Lambda (\frac{2}{\Lambda}-\frac{a}{\Lambda
		b}-\frac{b}{c})^\frac{1}{2}}]^\frac{1}{2},
\end{eqnarray}

where
\begin{equation}
a=3.{2}^\frac{1}{3}.(1-4Q^2 \Lambda),
\end{equation}

\begin{eqnarray}
b &=&[54-972 M^2 \Lambda+648 Q^2 \Lambda \nonumber\\
  & &+[(54-972 M^2 \Lambda+648 Q^2 \Lambda)^{2} \nonumber\\
  & &-4(9-36 Q^2 \Lambda)^3]^\frac{1}{2}]^\frac{1}{3},
\end{eqnarray}

\begin{equation}
c=3.{2}^\frac{1}{3} \Lambda,
\end{equation}

From the above equations we observe that $a$ depends on the charge of the black hole and the cosmological constant,
$b$ depends on the mass, the charge of the black hole and the cosmological constant whereas $c$ depends only on the cosmological constant.\\

{ We find from above that $r_3$ is negative and hence unphysical. $r_1, r_2, r_4$ are real and positive, depending upon suitable choice of $a, b, c, M, Q, \Lambda$. { Hence there are three possible horizons, from the innermost (depending on the values of $r$), they are Cauchy horizon, event horizon and the cosmological horizon.}\\

We get the values of the horizons for Reissner Nordstr\"{o}m metric, if we put
$\Lambda=0$ in eq.(4) above. Hence the values of
the radius of the horizon of the charged black hole in case of RN metric is,

\begin{equation}
r_{\pm} = M\pm{\sqrt{M^2-Q^2}},
\end{equation}

The larger one $r_{+}$, is the event horizon, while the smaller one, $r_{-}$,
is the inner or Cauchy horizon located inside the black hole.
The event horizon corresponds to,

\begin{equation}
r_{+} = M+{\sqrt{M^2-Q^2}},
\end{equation}
This is analog of the Schwarzschild radius, and for $Q=0$, $r_{+}=r_{s}=2M$.

\subsection{Solutions of Einstein-Maxwell equations in RN+$\Lambda$ spacetime:}

The metric (\ref{rnlmetric})[considering spherical and planar (2+1)-dimensional black holes as \cite{g}, \cite{h}]
can be written in the form,
\begin{equation}
ds^2 = -e^{\nu(r)} dt^2 + e^{\lambda(r)} dr^2 + r^2d\theta^2,
\end{equation}

where we take,
\begin{equation}
e^{\nu(r)}= e^{-\lambda(r)}=
(1-\frac{2M}{r}+\frac{Q^2}{r^2}-\frac{{\Lambda}r^2}{3}),
\end{equation}

The Hilbert action coupled to electromagnetism is given by ,
\begin{equation}
I = \int d x^3 \sqrt{-g } \left( \frac{R-2\Lambda}{16 \pi}
-\frac{1}{4} F_a^c F_{bc} + L_{m} \right),
\end{equation}

where $L_{m}$ is the Lagrangian for matter. The variation with respect to the
metric gives the following self consistent Einstein-Maxwell equations with
cosmological constant $\Lambda$ for a charged { cosmological constant effective fluid} distribution,

\begin{equation}
G_{ab}=R_{ab} - \frac{1}{2} R g_{ab} +\Lambda g_{ab} = - 8 \pi
(T_{ab}^{PF} +T_{ab}^{EM}),
\end{equation}

The explicit forms of the energy momentum tensor (EMT) components
for the matter source (we assumed that the matter distribution at
the interior of the black hole is { cosmological constant effective fluid} type) and
electromagnetic fields are given by,

\begin{equation}
T_{ab}^{PF} = (\rho +p) u_au_b + p g_{ab},
\end{equation}

\begin{equation}
T_{ab}^{EM} = -\frac{1}{4 \pi } \left( F_a^c F_{bc} -\frac{1}{4}
g_{ab} F_{cd}F^{cd}\right),
\end{equation}

where $\rho$, $p$, $u_i$ and $F_{ab}$ are, respectively,
matter density, fluid pressure and velocity three vector of
a fluid element and electromagnetic field. Here, the
electromagnetic field is related to current three vector

\begin{equation}
J^c = \sigma(r) u^c,
\end{equation}
as
\begin{equation}
F^{ab}_{;b} = - 4 \pi J^a,
\end{equation}
where, $\sigma(r) $ is the proper charge density of the
distribution. In our consideration, the three velocity is assumed
as $u_a = \delta_a^t$ and consequently the electromagnetic field tensor can be
given as,

\begin{equation}
F_{ab}  =  E(r) (\delta_a^t \delta_b^r-\delta_a^r \delta_b^t),
\end{equation}
where $E(r)$ is the electric field.\\

The Einstein-Maxwell equations with the assumption, cosmological constant
($\Lambda < 0$), for the black hole metric in eqn.(14) together
with the energy-momentum tensor given in
eqns. (18),(19) along with eqns. (20),(21) and (22) yield (rendering $G = c = 1$)

\begin{eqnarray}
\frac{\lambda' e^{-\lambda}}{2r} &=& 8\pi \rho(r) + E^2(r) +\Lambda
\label{eq9} \\ \frac{\nu' e^{-\lambda}}{2r} &=& 8\pi p(r) -E^2(r)
-\Lambda \label{eq10}  \\ \frac{e^{-\lambda}}{2}\left(\frac{1}{2}
\nu'^2+\nu''-\frac{1}{2}\nu'\lambda'\right) &=& 8\pi p(r)
+E^2(r) -\Lambda \label{eq11}\\
\sigma(r) =\frac{e^{-\frac{\lambda}{2}}}{4 \pi r^2 } (r^2 E(r))'\label{eq12}
\end{eqnarray}

where a `$\prime$' denotes differentiation with respect to the
radial parameter $r$. When E=0, the Einstein-Maxwell system given
above reduces to the uncharged Einstein system.

%

Here the term $\sigma(r)e^{\frac{\lambda(r)}{2}}$ \cite{m} is equivalent
to the volume charge density in ($2+1$)-d. We consider the proper charge
density $\sigma(r)$ as a polynomial function of $r$.\\
%
%
%
%
%

The E-M equations in (23) (24) and (25), using eqn.(15) reduce to,
\begin{equation}
8\pi \rho(r) +E^2(r) +\Lambda =
-\frac{1}{r}(\frac{M}{r^2}-\frac{Q^2}{r^3}-\frac{\Lambda r}{3})
\end{equation}

\begin{equation}
8\pi p(r) -E^2(r) -\Lambda = \frac{1}{r}(\frac{M}{r^2}-\frac{Q^2}{r^3}-\frac{\Lambda
r}{3})
\end{equation}

\begin{equation}
8\pi p(r) +E^2(r) -\Lambda = \frac{3 Q^2}{r^4}- \frac{2M}{r^3} -\frac{\Lambda}{3}
\end{equation}

%
%
%

On adding eqns. (27) and (28) we obtain,
\begin{eqnarray}
16\pi {p(r)} &=&-16\pi {\rho(r)} \nonumber\\
          &=&\frac{2Q^2}{r^4}-\frac{M}{r^3}+\frac{4\Lambda}{3}
\end{eqnarray}

The equations for pressure and matter density are evident from eqn.(30). Both should be dependent on the radial parameter r. It is observed from the above eqn.(30), that the particular case of $(\Lambda=0, r^2<\frac{2Q^2}{M})$ is not valid, which makes the energy density negative. Hence the cosmological constant should be negative, which is further verified in the following sections. The electric field E(r)
which is also dependent on r but is independent of the cosmological constant is given by
\begin{equation}
E(r)=(\frac{2Q^2}{r^4}-\frac{3M}{2r^3})^{\frac{1}{2}}
\end{equation}

It is observed from the above eqn.(31), that the electric field is non-vanishing and imaginary, if $Q=0$, and hence the field is limited to our $RN+\Lambda$ black hole, and cannot be reduced to the Schwarzschild form, as a particular case.

The proper charge density is also dependent on r, and from eqn.(26)is evaluated
as
\begin{equation}
\sigma(r)=\frac{(1-\frac{2M}{r}+\frac{Q^2}{r^2}-\frac{\Lambda
r^2}{3})^{\frac{1}{2}}(\frac{3M}{2r^3}-\frac{3Q^2}{r^4})}{2 {\pi} r
(\frac{2Q^2}{r^4}-\frac{3M}{2r^3})^\frac{1}{2}}
\end{equation}
\subsection{Physical significance of pressure and matter density:}
%
%
%

Thus for interior solutions we have deduced that $p=-\rho$. It is equivalent to
$p={\omega}\rho$, where we take ${\omega}=-1$.
This type of equation of state is available in the literature
and is known as a false vacuum, degenerate vacuum, or
$\rho$-vacuum and represents a repulsive pressure.\\

We choose the following values of the parameters,
\beq
\Lambda = -10^{-46} {\rm km^{-2}}, \quad M = 3.8 M_{\odot}, \quad Q= 0.00089 {\rm
km},
\eeq

The figures in the next page show the variation of $p(r)$ and $\rho(r)$ against
$r$.\\

\begin{figure}[htbp]
    \centering
        \includegraphics[scale=.85]{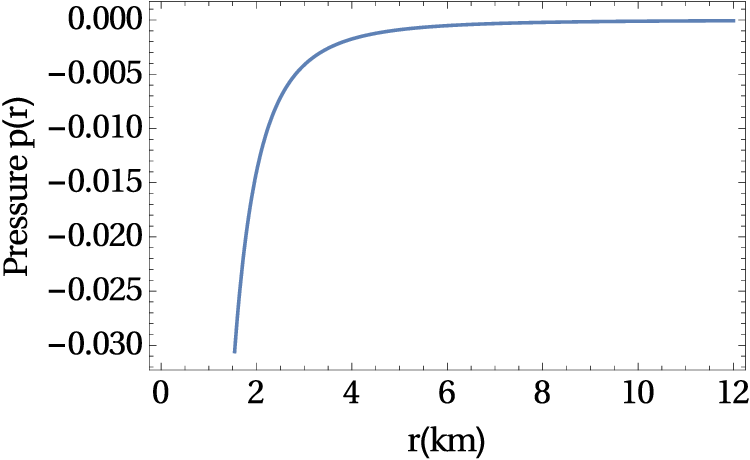}
       \caption{ Pressure $p(r)$ has been depicted against $r$. The geometric unit of pressure here is in $km^{-2}$.}
    \label{fig:1}
\end{figure}

\begin{figure}[htbp]
    \centering
        \includegraphics[scale=.85]{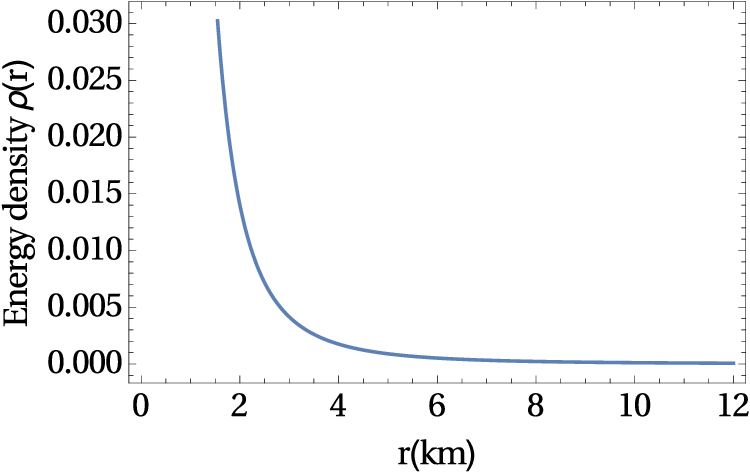}
       \caption{ Density $\rho(r)$ has been depicted against $r$. The geometric unit of density here is $km^{-2}$.}
    \label{fig:2}
\end{figure}

\begin{figure}[htbp]
    \centering
        \includegraphics[scale=.6]{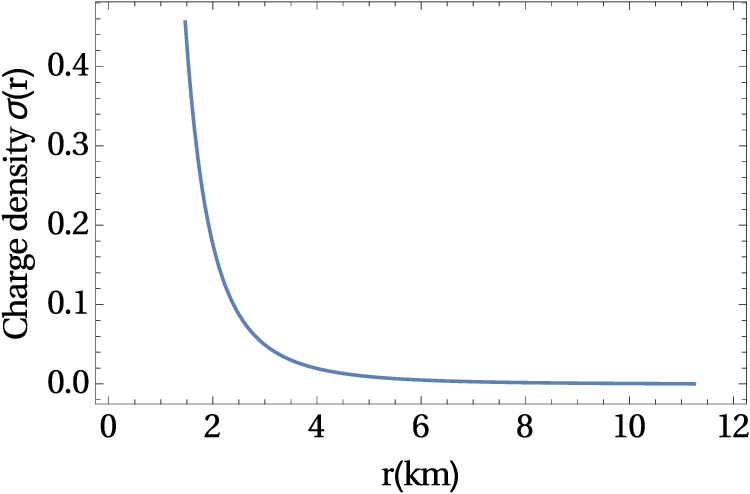}
       \caption{Proper charge density $\sigma(r)$ has been depicted against $r$. The geometric unit of proper charge density here is $km^{-2}$.}
    \label{fig:3}
\end{figure}

Hence we observe from the figure, that the black hole has a negative pressure and positive matter density inside which is due to the presence of exotic matter. { The pressure $p(r)$ increases with the increase in radius and the matter density $\rho(r)$ decreases with the increase in radius}.\\

{ We have assumed that the effective mass of the black hole is $M = 3.8 M_{\odot}$}. It can also be observed via
eqn.(30), that the physical parameters, viz.
density and pressure are dependent on the charge. Also, if $\sigma(r)=0$, then from eqn.(32) we get $M=\frac{2Q^2}{r}$, and both the parameters $p(r)$ and $\rho(r)$ in eqn.(30), become constant, being dependent only on $\Lambda$. Therefore our solutions provide {\it electromagnetic mass} model, such that for vanishing charge density $\sigma(r)$, the physical parameters (pressure and density) becomes constant.\\

Figure 3. shows the variation of the proper charge density against $r$. We observe
that the proper charge density is maximum at the centre and decreases with the increase in radius.

\subsection{Geodesic equations in RN+$\Lambda$ spacetime:}

We write the geodesic equations as follows:-
\begin{equation}
\frac{d^{2}t}{ds^2}+ \nu' \frac{dt}{ds} \frac{dr}{ds} = 0,
\end{equation}

\begin{equation}
\frac{d^{2}r}{ds^2}+ \frac{1}{2}{\nu'}e^{(\nu-\lambda)}(\frac{dt}{ds})^2
+\frac{1}{2}\lambda'(\frac{dr}{ds})^2 -re^{-\lambda} (\frac{d\theta}{ds})^2 = 0,
\end{equation}

\begin{equation}
\frac{d^{2}\theta}{ds^2}+ \frac{2}{r} \frac{d\theta}{ds} \frac{dr}{ds}= 0,
\end{equation}

Using eqn.(15), since $\nu =-\lambda$, $\nu'=-\lambda'$ and $\nu-\lambda=2\nu$,
we find,
\begin{equation}
\nu'= \frac{2}{(1-\frac{2M}{r}+\frac{Q^2}{r^2}-\frac{\Lambda
r^2}{3})}.(\frac{M}{r^2}-\frac{Q^2}{r^3}-\frac{\Lambda r}{3}),
\end{equation}

On integrating eqns.(34) and (36), we obtain
\begin{equation}
\frac{d \theta}{ds} = \frac{k_{1}^2}{r^2},
\end{equation}

and

\begin{equation}
\frac{d t}{ds} = \frac{k_{2}^2}{(1-\frac{2M}{r}+\frac{Q^2}{r^2}-\frac{\Lambda
r^2}{3})},
\end{equation}

Putting the above values of $\frac{d \theta}{ds}$ and $\frac{d t}{ds}$ in
eqn.(35), we get,
\begin{eqnarray}
\frac{d^{2}r}{ds^2}+ \frac{k_{2}^2(\frac{M}{r^2}-\frac{Q^2}{r^3}-\frac{\Lambda
r}{3})}{(1-\frac{2M}{r}+\frac{Q^2}{r^2}-\frac{\Lambda r^2}{3})}\nonumber\\
-\frac{(\frac{M}{r^2}-\frac{Q^2}{r^3}-\frac{\Lambda
r}{3})}{(1-\frac{2M}{r}+\frac{Q^2}{r^2}-\frac{\Lambda r^2}{3})}
.(\frac{dr}{ds})^2\nonumber\\-\frac{k_{1}^2}{r^3}.(1-\frac{2M}{r}+\frac{Q^2}{r^2}-\frac{
\Lambda r^2}{3})=0,
\end{eqnarray}

Also from the metric eqn.(14), on dividing each side by $ds^2$, we find that,
\begin{equation}
1 = -e^\nu (\frac{dt}{ds})^2+ e^{\lambda}(\frac{dr}{ds})^2+ r^2 (\frac{d
\theta}{ds})^2,
\end{equation}

Using eqns.(38) and (39), eqn.(41) reduces to,
\begin{equation}
\Big(\frac{d r}{ds}\Big)^{2} = k_{2}^2 +\Big(1-\frac{2M}{r}+\frac{Q^2}{r^2}-\frac{\Lambda
r^2}{3}\Big)\Big(1-\frac{k_{1}^2}{r^2}\Big),
\end{equation}

Using eqn.(42) in eqn.(40), we obtain
\begin{eqnarray}
\frac{d^2 r}{d s^2}&=&\Big(\frac{M}{r^2}-\frac{Q^2}{r^3}-\frac{\Lambda
r}{3}\Big)\nonumber\\
                   & &+ k_{1}^2 \Big(\frac{1}{r^3}-\frac{3M}{r^4}+\frac{2Q^2}{r^5}\Big),
\end{eqnarray}

Multiplying the above equation by $2\frac{dr}{ds}$ and integrating both sides
w.r.t $ds$ we get,
\begin{equation}
\Big(\frac{d r}{ds}\Big)^{2} = 2[-\frac{M}{r}+\frac{Q^2}{2r^2}-\frac{{\Lambda}r^2}{6}+
k_{1}^2 \Big(-\frac{1}{2r^2}+\frac{M}{r^3}-\frac{Q^2}{2r^4}\Big)],
\end{equation}

On equating eqns.(42) and (44), we observe that,
\begin{equation}
k_{2}^2=-\Big(\frac{\Lambda k_{1}^2}{3}+ 1\Big),
\end{equation}

Hence we observe that the cosmological constant can have a negative value
as is confirmed from the above eqn.(45) \cite{maeda14}.
Hence our assumption is found to be true.

\section{Metric for interior RN+$\Lambda$ and exterior FRW
spacetimes:}\label{sec3}

The metric for RN+$\Lambda$ black hole in $(2+1)$ dimensions is given by
eqn.(1). For the
sake of convenience we transform the metric under isotropic conditions with the
following transformations, using $x^0=v$ and $x^1=x$, as

\begin{equation}
2t=v, 2s=l, 2r=x[(1+\frac{M}{x})^2-\frac{Q^2}{x^2}+\Lambda e^{-2x}],
\end{equation} \label{eqn.46}

Hence eqn.(1) is transformed as,
\begin{eqnarray}
dl^{2}&=&-\frac{(1-\frac{M^2}{x^2}+\frac{Q^2}{x^2}-\Lambda
e^{-2x})^2}{[(1+\frac{M}{x})^2-\frac{Q^2}{x^2}+\Lambda e^{-2x}]^2} dv^{2}
\nonumber\\
      & &+ [(1+\frac{M}{x})^2-\frac{Q^2}{x^2}+\Lambda e^{-2x}]^2 \nonumber\\
      & &.(dx^2+x^{2} d\theta^{2}), \label{eqn.47}
\end{eqnarray}

We consider the line element for the exterior space-time in FRW metric in the
form,
\begin{eqnarray}
dl^{2}&=&-dv^{2} + \frac{a^{2}(v)}{(1+\frac{kx^2}{4})^2} 
     (dx^2+x^{2} d\theta^{2}), \label{eqn.48}
\end{eqnarray}

Here $a(v)$ is the scale factor of the universe and k denotes the space-time
curvature.
The above RN+$\Lambda$ metric embedded in FRW universe is represented as
follows:

\begin{eqnarray}
dl^{2}&=&- P^{2}(v,x) dv^{2} + T^{2}(v,x) (dx^2+x^{2} d\theta^{2}),
\end{eqnarray} \label{eqn.49}

where,
\begin{eqnarray}
T(v,x)&=&[g(v,x)+\frac{z(v)}{x}]^{2}-\frac{h(v)}{x^2}\nonumber\\
      & &+\Lambda g^{2} e^{-2 b g x}, \label{eqn.50}
\end{eqnarray}

\begin{eqnarray}
P(v,x)&=&f(v) \frac{\dot{T}}{2T} \nonumber\\
      &=& [\frac{\dot{g}f}{g}+(g \dot{z}+\dot{g} z)\frac{f}{g^{2} x}+\frac{z
\dot{z} f}{g^{2} x^{2}}- \frac{\dot{h} f}{2 g^{2} x^{2}}\nonumber\\
      & &-\Lambda e^{-2bgx}].[(1+\frac{z}{g
x})^{2}-\frac{h}{g^{2} x^{2}}\nonumber\\
      & &+\Lambda e^{-2bgx}]^{-1}, \label{eqn.51}
\end{eqnarray}

where ``.'' denotes differentiation with respect to $v$ and $b(v)$ is the scale factor under the transformed conditions. We consider the limit when $f \dot{g}(b x-\frac{1}{g})\rightarrow1$.\\

We consider asymptotic flat conditions where $P(v=const.,x)$ is reduced to
$\sqrt{g_{00}}$ term in eqn.(47). Hence on comparing the $\sqrt{g_{00}}$ term of eqn.(51) with that of (47), we find that the following identities hold:(i)$\frac{\dot{g}f}{g}=1$, (ii) $(g \dot{z}+ \dot{g} z )\frac{f}{g^{2}x}=0$, (iii) $\frac{z \dot{z} f }{g^2 x^2}=-(\frac{z}{g x})^2 $ and (iv)$-\frac{\dot{h} f}{2 g^2 x^2}=\frac{h}{g^2 x^2}$.\\

Now, (i)-(iv) reduce to (v) $\dot{g}f=g, \dot{z} f=-z, \dot{h} f=-2h$;\\

On suitable transformations assuming, $f=\frac{b}{\dot{b}}$ and $g=\frac{b(v)}{\sqrt{1+\frac{k x^2}{4}}}$, we obtain,
(vi) $z=\frac{M}{b}$, $h=\frac{Q^2}{b^2}$.\\

Here the integration constants $M$ and $Q$ are related to the mass and charge of the
black
hole respectively.\\

On substituting the above eqns.(49), (50) and (51) in (47) with the above transformations under suitable conditions, the final
RN+$\Lambda$ metric in $(2+1)$ dimensions in the FRW background
is observed as follows:

\begin{eqnarray}
dl^{2}&=&-\frac{(1-\frac{M^2
(1+\frac{kx^2}{4})}{a^{2}x^{2}}+\frac{Q^{2}(1+\frac{kx^2}{4})}{a^{2}x^2}
-\Lambda e^{-\frac{2ax}{\sqrt{1+\frac{kx^2}{4}}}})^2}
{[(1+\frac{M {\sqrt{1+\frac{kx^2}{4}}}}{ax})^2-\frac{Q^2
(1+\frac{kx^2}{4})}{a^{2}x^2}+\Lambda
e^{-\frac{2ax}{\sqrt{1+\frac{kx^2}{4}}}}]^2} \nonumber\\
      & &.dv^{2} + \frac{a^2}{(1+\frac{kx^2}{4})^2}.[(1+\frac{M
{\sqrt{1+\frac{kx^2}{4}}}}{ax})^2 \nonumber\\
      & &-\frac{Q^2 (1+\frac{kx^2}{4})}{a^{2}x^2}+\Lambda
e^{-\frac{2ax}{\sqrt{1+\frac{kx^2}{4}}}}]^2 \nonumber\\
      & &.(dx^2+x^{2} d\theta^{2}), \label{eqn.52}
\end{eqnarray}

Here $a=a(v)$ $\rightarrow$ $b^{2}(v)$.\\

If k=0, the above eqn.(52) reduces to,

\begin{eqnarray}
dl^{2}&=&-\frac{(1-\frac{M^2}{a^{2}x^{2}}+\frac{Q^{2}}{a^{2}x^2}-\Lambda
e^{-2ax})^2}
{[(1+\frac{M}{ax})^2-\frac{Q^2}{a^{2}x^2}+\Lambda e^{-2ax}]^2} dv^{2}
\nonumber\\
      & &+ {a^2} [(1+\frac{M}{ax})^2 -\frac{Q^2}{a^{2}x^2}+\Lambda e^{-2ax}]^2
\nonumber\\
      & &.(dx^2+x^{2} d\theta^{2}),
\end{eqnarray}

We know that $a(v)=e^{Hv}$ where H is the Hubble constant. If further $H=0$, then
$a(v)=1$ and the eqn.(47) is restored from eqn.(53). However $Q=0$
reduces the above eqn.(52) to,

\begin{eqnarray}
dl^{2}&=&-\frac{(1-\frac{M^2 (1+\frac{kx^2}{4})}{a^{2}x^{2}}-\Lambda
e^{-\frac{2ax}{\sqrt{1+\frac{kx^2}{4}}}})^2}
{[(1+\frac{M {\sqrt{1+\frac{kx^2}{4}}}}{ax})^2+\Lambda
e^{-\frac{2ax}{\sqrt{1+\frac{kx^2}{4}}}}]^2} dv^{2} \nonumber\\
      & &+ \frac{a^2}{(1+\frac{kx^2}{4})^2}.[(1+\frac{M
{\sqrt{1+\frac{kx^2}{4}}}}{ax})^2 \nonumber\\
      & &+\Lambda e^{-\frac{2ax}{\sqrt{1+\frac{kx^2}{4}}}}]^2.(dx^2+x^{2}
d\theta^{2}),
\end{eqnarray}

which is just the McVittie solution. For the extreme RN black hole case $M=Q$
and $k=0$ in presence of cosmological constant,
the eqn.(52) is reduced to

\begin{eqnarray}
dl^{2}&=&-\frac{(1-\Lambda e^{-2ax})^2}
{(1+\frac{2M}{ax}+\Lambda e^{-2ax})^{2}} dv^{2} \nonumber\\
      & &+ {a^2} (1+\frac{2M}{ax}+\Lambda e^{-2ax})^{2} \nonumber\\
      & &.(dx^2+x^{2} d\theta^{2}),
\end{eqnarray}

If $\Lambda=0$ the above eqn.(55) further takes the form,

\begin{eqnarray}
dl^{2}&=&-\frac{1}{(1+\frac{2M}{ax})^2} dv^{2}+ {a^2}(1+\frac{2M}{ax})^{2}
\nonumber\\
      & &.(dx^2+x^{2} d\theta^{2}),
\end{eqnarray}

If the scale factor $a(v)=1$, when $H=0$, the above eqn.(56) reduces to the
Schwarzschild
metric in an FRW present day accelerating universe as,

\begin{eqnarray}
dl^{2}&=&-\frac{1}{(1+\frac{2M}{x})^2} dv^{2} \nonumber\\
      & &+ (1+\frac{2M}{x})^{2} (dx^2+x^{2} d\theta^{2}),
\end{eqnarray}

\section{Boundary and matching conditions with the exterior FRW universe:}\label{sec4}

We use matching conditions of $g_{vv}$, $g_{xx}$ and $\frac{\partial
g_{vv}}{\partial x}$ at $x=R$
we find from eqns. (48) and (52) three resuls which are enunciated below,

\subsection{Continuity of $g_{vv}$:}
\begin{equation}
\frac{(1-\frac{M^2
(1+\frac{kx^2}{4})}{a^{2}x^{2}}+\frac{Q^{2}(1+\frac{kx^2}{4})}{a^{2}x^2}
-\Lambda e^{-\frac{2ax}{\sqrt{1+\frac{kx^2}{4}}}})^2}
{[(1+\frac{M {\sqrt{1+\frac{kx^2}{4}}}}{ax})^2-\frac{Q^2
(1+\frac{kx^2}{4})}{a^{2}x^2}+\Lambda
e^{-\frac{2ax}{\sqrt{1+\frac{kx^2}{4}}}}]^2}=1
\end{equation}

Thus the scale factor $a(v)$ is expressed by the following equation,

\begin{eqnarray}
\frac{M^2 (1+\frac{kx^2}{4})}{a^{2}x^{2}} +\frac{M {\sqrt{1+\frac{kx^2}{4}}}}{ax}-\frac{Q^{2}(1+\frac{kx^2}{4})}{a^{2}x^2}\nonumber\\
+\Lambda e^{-\frac{2ax}{\sqrt{1+\frac{kx^2}{4}}}}=0,
\end{eqnarray}

Hence $\Lambda$ is negative for a positive mass.
For the extreme R-N case when $Q=M$,

\begin{eqnarray}
\frac{M {\sqrt{1+\frac{kx^2}{4}}}}{ax}+\Lambda e^{-\frac{2ax}{\sqrt{1+\frac{kx^2}{4}}}} \nonumber\\
 =\frac{Q {\sqrt{1+\frac{kx^2}{4}}}}{ax}+\Lambda e^{-\frac{2ax}{\sqrt{1+\frac{kx^2}{4}}}}=0, \nonumber\\
\end{eqnarray}

As $\Lambda$ is constant the above eqn.(60) indicates that, for an observer at
infinity,
the mass and charge of the black hole decreases with the expansion of the
universe whereas both increases with the
contraction of the universe.\\

If $Q=0$ we get,
\begin{eqnarray}
[\frac{M {\sqrt{1+\frac{kx^2}{4}}}}{ax}].[1+\frac{M {\sqrt{1+\frac{kx^2}{4}}}}{ax}]+\Lambda e^{-\frac{2ax}{\sqrt{1+\frac{kx^2}{4}}}}\nonumber\\
=0,
\end{eqnarray}

\subsection{Continuity of $g_{xx}$:}
\begin{eqnarray}
\frac{a^2(v)}{(1+\frac{kx^2}{4})^2}&=&\frac{a^2(v)}{(1+\frac{kx^2}{4})^2}.[
(1+\frac{M {\sqrt{1+\frac{kx^2}{4}}}}{ax})^2 \nonumber\\
      & &-\frac{Q^2 (1+\frac{kx^2}{4})}{a^{2}x^2}+\Lambda
e^{-\frac{2ax}{\sqrt{1+\frac{kx^2}{4}}}}]^2
\end{eqnarray}

The above eqn.(62) gives another expression for $a(v)$ as,
\begin{eqnarray}
\frac{M^2 (1+\frac{kx^2}{4})}{a^{2}x^{2}} +\frac{2M {\sqrt{1+\frac{kx^2}{4}}}}{ax}-\frac{Q^{2}(1+\frac{kx^2}{4})}{a^{2}x^2}\nonumber\\
+\Lambda e^{-\frac{2ax}{\sqrt{1+\frac{kx^2}{4}}}}=0,
\end{eqnarray}

We observe via eqns.(59) and (63) that as $\Lambda$ has a negative value of the order $10^{-46}$,
the term containing $\Lambda$ can be eliminated to obtain the value of the constant $k=-0.0278$ in both the equations, considering
the scale factor, $a(v)=1$ for the present day accelerating universe.

\subsection{Continuity of $\frac{\partial g_{vv}}{\partial x}$ at $x=R$:}

\begin{eqnarray}
2(2-\frac{\sqrt{1+\frac{kx^2}{4}}}{ax})[\frac{M^2}{(\frac{ax}{\sqrt{1+\frac{kx^2
}{4}}})^3}
-\frac{Q^2}{(\frac{ax}{\sqrt{1+\frac{kx^2}{4}}})^3} \nonumber\\
+\Lambda e^{-\frac{2ax}{\sqrt{1+\frac{kx^2}{4}}}}]+\frac{\sqrt{1+\frac
{kx^2}{4}}}{ax}=0,
\end{eqnarray}

Hence at $x=R$ we get,
\begin{eqnarray}
2(2-\frac{\sqrt{1+\frac{kR^2}{4}}}{a R})[\frac{M^2}{(\frac{a
R}{\sqrt{1+\frac{kR^2}{4}}})^3}
-\frac{Q^2}{(\frac{a R}{\sqrt{1+\frac{kR^2}{4}}})^3} \nonumber\\
+\Lambda e^{-\frac{2a R}{\sqrt{1+\frac{kR^2}{4}}}}]+\frac{\sqrt{1+\frac
{kR^2}{4}}}{a R}=0,
\end{eqnarray}

In the extreme R-N case when $Q=M$ we find,

\begin{equation}
\Lambda = - e^{\frac{2a R}{\sqrt{1+\frac{kR^2}{4}}}}.\frac{1}{2(\frac{2a R}{{\sqrt{1+\frac{kR^2}{4}}}}-1)}
\end{equation}

As $\Lambda$ is negative,$\frac{2a R}{{\sqrt{1+\frac{kR^2}{4}}}}{>}1$.
Hence $R {>}\frac{2}{\sqrt{16a^{2}-k}}$,
where k is the curvature of space-time. The curvature parameter k may take values of 0, +1 or -1, depending on whether 3-D spacetime is assumed to be Euclidean, spherical, or hyperbolic, respectively. Here we observe that R is always positive for the accelerating universe, if we take $a(v)=1$ and $k=1$.\\

For $Q=0$ in eqn.(64),
\begin{eqnarray}
\Lambda = - e^{\frac{2a R}{\sqrt{1+\frac{kR^2}{4}}}}.[\frac{M^2}{(\frac{a
R}{\sqrt{1+\frac{kR^2}{4}}})^3} \nonumber\\
+\frac{1}{2(\frac{2a R}{{\sqrt{1+\frac{kR^2}{4}}}}-1)}]
\end{eqnarray}
which represents the cosmological constant inside the Schwarzschild black hole and also has
a negative value.

We have considered $Q=0.00089~km$ and c=1, as geometric units. However when converted to SI units we get $Q=1.03419 \times 10^{17}~coulomb$.

\section{Physical relevance of RN+$\Lambda$ metric:}\label{sec5}

It is found that the final RN+$\Lambda$ metric in eqn.(52) satisfies the field equations (23)-(26). The Einstein tensor $G_{\mu \nu}$ and the energy momentum tensor $T_{ab}^{PF}$ and $T_{ab}^{EM}$ for { cosmological constant effective fluid} and electromagnetic fields w.r.t the metric are easily obtained. Using eqns. (18)-(22) we deduce the non-vanishing components of the electromagnetic tensor $F_{ab}$ as,

\begin{eqnarray}
F^{01}=E^{1}=[2(4+kx^2)(e^{\frac{4ax}{\sqrt{4+kx^2}}}(-4a^2 x^2 + M^2\nonumber\\ (4+kx^2)-Q^2 (4+kx^2))+4a^2x^2 \Lambda)(e^{\frac{4ax}{\sqrt{4+kx^2}}}\nonumber\\(4a^2x^2+4aMx \sqrt{4+kx^2}+M^2(4+kx^2)\nonumber\\-Q^2 (4+kx^2))+4a^2x^2 \Lambda)^4]^{-\frac{1}{2}}\times[32a^3 e^{\frac{12ax}{\sqrt{4+kx^2}}}x^3\nonumber\\ ((4+kx^2)(e^\frac{8ax}{\sqrt{4+kx^2}}M(M^2-Q^2)^2  (-2+kx^2)\nonumber\\(4+kx^2)^\frac{5}{2}+4ae^\frac{8ax}{\sqrt{4+kx^2}}(M^2-Q^2)x(4+kx^2)^2 \nonumber\\(2k M^2 x^2 +Q^2 (4-kx^2))+8a^2 e^\frac{4ax}{\sqrt{4+kx^2}}\nonumber\\ M(M^2-Q^2)x^2 (4+kx^2)^{\frac{3}{2}}(3 e^\frac{4ax}{\sqrt{4+kx^2}}\nonumber\\ (2+kx^2)+8 \Lambda)+16a^3 e^\frac{4ax}{\sqrt{4+kx^2}} x^3(4+kx^2)\nonumber\\(e^\frac{4ax}{\sqrt{4+kx^2}}(2M^2 (4+kx^2)-Q^2 (8+kx^2))\nonumber\\+(8M^2-Q^2 (8+kx^2))\Lambda)+16a^4 Mx^4 \sqrt{4+kx^2}\nonumber\\(6+kx^2)(e^\frac{8ax}{\sqrt{4+kx^2}}-\Lambda^2) )+16a^3 x^3 \nonumber\\\Lambda (e^\frac{4ax}{\sqrt{4+kx^2}}(M^3 (-16+kx^2)(4+kx^2)^2\nonumber\\ +6a M^2 x {\sqrt{4+kx^2}} (-40-6kx^2+k^2 x^4)+\nonumber\\2ax {\sqrt{4+kx^2}} (4a^2 x^2 (2+kx^2)+Q^2 (88+18kx^2\nonumber\\-k^2 x^4))-M (4+kx^2)(-12 a^2 x^2 (-4+kx^2)\nonumber\\+Q^2 (-64-12kx^2+k^2 x^4)))+4a^2 x^2\nonumber\\ (2ax (2+kx^2) {\sqrt{4+kx^2}}+M (32+12 kx^2\nonumber\\ +k^2 x^4)) \Lambda )+128 a^4 x^4 (2a e^\frac{4ax}{\sqrt{4+kx^2}} (3M^2\nonumber\\-Q^2)x(4+kx^2)+e^\frac{4ax}{\sqrt{4+kx^2}}M(M^2-Q^2)\nonumber\\(4+kx^2)^{\frac{3}{2}}+8a^3 x^3 (e^\frac{4ax}{\sqrt{4+kx^2}}-2\Lambda)+4a^2 M x^2\nonumber\\ {\sqrt{4+kx^2}}(3 e^\frac{4ax}{\sqrt{4+kx^2}}-2\Lambda))\Lambda ]^{\frac{1}{2}}
\end{eqnarray}

which reduces significantly as follows when $Q=M$,
\begin{eqnarray}
F^{01}=E^{1}=[2(4+kx^2)(e^{\frac{4ax}{\sqrt{4+kx^2}}}(-4a^2 x^2  )+4a^2x^2 \Lambda)\nonumber\\(e^{\frac{4ax}{\sqrt{4+kx^2}}}(4a^2x^2+4aMx \sqrt{4+kx^2})+4a^2x^2 \Lambda)^4]^{-\frac{1}{2}}\nonumber\\\times[32a^3 e^{\frac{12ax}{\sqrt{4+kx^2}}}x^3((4+kx^2)(16a^3 e^\frac{4ax}{\sqrt{4+kx^2}} x^3 (4+kx^2)\nonumber\\(e^\frac{4ax}{\sqrt{4+kx^2}}(kM^2 x^2)-kM^2 x^2\Lambda)+16a^4 Mx^4 \sqrt{4+kx^2}\nonumber\\(6+kx^2)(e^\frac{8ax}{\sqrt{4+kx^2}}-\Lambda^2) +16a^3 x^3 \Lambda (e^\frac{4ax}{\sqrt{4+kx^2}}\nonumber\\(M^3 (-16+kx^2)(4+kx^2)^2+6a M^2 x {\sqrt{4+kx^2}}\nonumber\\ (-40-6kx^2+k^2 x^4)+2ax {\sqrt{4+kx^2}} (4a^2 x^2\nonumber\\ (2+kx^2)+M^2 (88+18kx^2-k^2 x^4))-M (4+kx^2)\nonumber\\(-12 a^2 x^2 (-4+kx^2)+M^2 (-64-12kx^2+k^2 x^4)))\nonumber\\+4a^2 x^2(2ax (2+kx^2) {\sqrt{4+kx^2}}+M (32+12 kx^2\nonumber\\ +k^2 x^4)) \Lambda )+128 a^4 x^4 (4a M^2 e^\frac{4ax}{\sqrt{4+kx^2}}\nonumber\\ x(4+kx^2)+8a^3 x^3 (e^\frac{4ax}{\sqrt{4+kx^2}}-2\Lambda)+4a^2 M x^2 \nonumber\\{\sqrt{4+kx^2}}(3 e^\frac{4ax}{\sqrt{4+kx^2}}-2\Lambda))\Lambda]^{\frac{1}{2}}~~~~~
\end{eqnarray}
Also since $F_{a b}=A_{a;b}-A_{b;a}$ , using eqn. (68), we get the non-vanishing components of the potential $A_{a}$ as,
\begin{eqnarray}
A_{0}=\int F^{01} g_{00} g_{11} dx
\end{eqnarray}

Furthermore the eqn.(68) satisfies $F^{a b}_{; b}=0$. From eqn.(17) $G^{a b}_{; b}=0$  always holds, hence we get,
$T_{ab;b}^{PF} +T_{ab;b}^{EM}=0$. We also find that the above relation is satisfied using equations (18) and (19) as,
$u_{a}T_{ab;b}^{PF}=T_{0b;b}^{PF}= p_{,b} g_{0b} + p g_{0b;b}+(\rho +p)_{,b}u_{b}u_{0}+(\rho +p)[u_{b;b}u_{0}+u_{0;b}u_{b}]=0$ and $4 \pi T_{;b}^{ab (EM)}=F^{a \alpha}_{;b}F^{b}_{\alpha}+F^{a \alpha}F^{b}_{\alpha;b}-\frac{1}{2}g^{ab}F_{\alpha \beta}F^{\alpha \beta}=F^{a \alpha}F^{b}_{\alpha;b}+\frac{1}{2}g^{a \zeta}F^{b \kappa}(F_{\zeta \kappa ;b}-F_{\zeta b ;\kappa}-F_{\kappa b ;\zeta})=-F^{a \alpha} J_{\alpha}=0$.\\

So both $T_{ab}^{PF}$ and $T_{ab}^{EM}$ satisfy Bianchi identity. The proof is indicative of the fact that eqn.(52) is an exact solution of the $EM$ field equations and the metric is physically relevant.

\section{Darmois-Israel matching conditions:}\label{sec6}

The Darmois-Israel matching conditions have been studied [\cite{aa},\cite{bb}]. The junction conditions to match the inner and exterior metrics across the boundary surface $x=x_{0}$, are the continuity of first and second fundamental forms across that surface. We define a surface $\sum$, where $x=x_{0}$, the junction surface being an one dimensional ring of matter, by the metric \cite{cc},
\begin{eqnarray}
dl^{2}_{\sum}=-d{\tau^2}+x^2_{0}d {\theta}^2,
\end{eqnarray}
with the intrinsic coordinates of $\sum$ being $\xi^m=(\tau,\theta)$. The inner and outer metrics from eqns. (52) and (48) are given as,
\begin{eqnarray}
dl^{2}_{-}&=&-\frac{(1-\frac{M^2
		(1+\frac{kx^2}{4})}{a^{2}x^{2}}+\frac{Q^{2}(1+\frac{kx^2}{4})}{a^{2}x^2}
	-\Lambda e^{-\frac{2ax}{\sqrt{1+\frac{kx^2}{4}}}})^2}
{[(1+\frac{M {\sqrt{1+\frac{kx^2}{4}}}}{ax})^2-\frac{Q^2
		(1+\frac{kx^2}{4})}{a^{2}x^2}+\Lambda
	e^{-\frac{2ax}{\sqrt{1+\frac{kx^2}{4}}}}]^2} \nonumber\\
& &.dv^{2} + \frac{a^2}{(1+\frac{kx^2}{4})^2}.[(1+\frac{M
	{\sqrt{1+\frac{kx^2}{4}}}}{ax})^2 \nonumber\\
& &-\frac{Q^2 (1+\frac{kx^2}{4})}{a^{2}x^2}+\Lambda
e^{-\frac{2ax}{\sqrt{1+\frac{kx^2}{4}}}}]^2 \nonumber\\
& &.(dx^2+x^{2} d\theta^{2}),
\end{eqnarray}
and
\begin{eqnarray}
dl^{2}_{+}&=&-dv^{2} + \frac{a^{2}(v)}{(1+\frac{kx^2}{4})^2} 
(dx^2+x^{2} d\theta^{2}),
\end{eqnarray}
Here the coordinates $(v,x,\theta)$ are recognised in both the regions of the spacetime.

Now we consider the boundary surface $\sum$ as timelike which would imply
\begin{eqnarray}
\frac{(1-\frac{M^2
		(1+\frac{kx^2}{4})}{a^{2}x^{2}}+\frac{Q^{2}(1+\frac{kx^2}{4})}{a^{2}x^2}
	-\Lambda e^{-\frac{2ax}{\sqrt{1+\frac{kx^2}{4}}}})^2}
{[(1+\frac{M {\sqrt{1+\frac{kx^2}{4}}}}{ax})^2-\frac{Q^2
		(1+\frac{kx^2}{4})}{a^{2}x^2}+\Lambda
	e^{-\frac{2ax}{\sqrt{1+\frac{kx^2}{4}}}}]^2}>0 \nonumber\\
\label{eqn.74}
\end{eqnarray}

The radial coordinate $x$ is used as the matching parameter along the generators on $\sum$, the normal $\eta_{m}$ to the surface has only the radial component $\eta_{x}=\sqrt{g_{xx}}$. We thus obtain the extrinsic curvature in the form $[where, x^{0}=v, x^{1}=x, x^{2}=\theta]$,
\begin{eqnarray}
K^{\pm}_{mn}=-{\eta^{\pm}_{x}} {\Gamma^{x(\pm)}_{ab}} \frac{\partial x^{a}}{\partial \xi^{m}} \frac{\partial x^{b}}{\partial \xi^{n}} ,
\end{eqnarray}

Now, the line elements in eqns.(72) and (73) are continuous at $x=x_{0}$. The continuity of the first fundamental form at the boundary indicates that $g_{vv}^{+}=g_{vv}^{-}$ and $g_{xx}^{+}=g_{xx}^{-}$, i.e,
\begin{equation}
\frac{(1-\frac{M^2
		(1+\frac{kx_{0}^2}{4})}{a^{2}x_{0}^{2}}+\frac{Q^{2}(1+\frac{kx_{0}^2}{4})}{a^{2}x_{0}^2}
	-\Lambda e^{-\frac{2ax_{0}}{\sqrt{1+\frac{kx_{0}^2}{4}}}})^2}
{[(1+\frac{M {\sqrt{1+\frac{kx_{0}^2}{4}}}}{ax_{0}})^2-\frac{Q^2
		(1+\frac{kx_{0}^2}{4})}{a^{2}x_{0}^2}+\Lambda
	e^{-\frac{2ax_{0}}{\sqrt{1+\frac{kx_{0}^2}{4}}}}]^2}=1
\end{equation}
and
\begin{eqnarray}
\frac{a^2(v)}{(1+\frac{kx_{0}^2}{4})^2}&=&\frac{a^2(v)}{(1+\frac{kx_{0}^2}{4})^2}.[
(1+\frac{M {\sqrt{1+\frac{kx_{0}^2}{4}}}}{ax_{0}})^2 \nonumber\\
& &-\frac{Q^2 (1+\frac{kx_{0}^2}{4})}{a^{2}x_{0}^2}+\Lambda
e^{-\frac{2ax_{0}}{\sqrt{1+\frac{kx_{0}^2}{4}}}}]^2
\end{eqnarray}
Hence we retrieve eqns.(58) and (62) on replacing $x_{0}$ by $x$ in the above eqns.(76) and (77) respectively. It is also evident that,
\begin{eqnarray}
K^{-}_{\tau \tau}=-{\eta^{-}_{x}} {\Gamma^{(-)x}_{vv}} \frac{dv}{d \tau} \frac{dv}{d \tau},
\end{eqnarray}
But $g_{vv}^{\pm} \frac{dv^2}{d \tau^2}=-1$ (by construction) and
\begin{eqnarray}
\eta_{x}^{-}=\frac{a}{(1+\frac{kx_{0}^2}{4})}[(1+\frac{M
	{\sqrt{1+\frac{kx_{0}^2}{4}}}}{ax_{0}})^2 \nonumber\\
-\frac{Q^2 (1+\frac{kx_{0}^2}{4})}{a^{2}x_{0}^2}+\Lambda
e^{-\frac{2ax_{0}}{\sqrt{1+\frac{kx_{0}^2}{4}}}}]
\end{eqnarray}
Now,
\begin{eqnarray}
\frac{\partial g_{vv}^{-}}{\partial x}=[(4+kx_{0}^2)^{-3/2}(e^{\frac{4ax_{0}}{\sqrt{4+kx_{0}^2}}}(4a^2 x_{0}^2\nonumber\\+4a Mx_{0} \sqrt{4+kx_{0}^2}+M^2 (4+kx_{0}^2)-Q^2\nonumber\\ (4+kx_{0}^2))+4a^2 x_{0}^2 \Lambda)^{-3}][32 ae^{\frac{4ax_{0}}{\sqrt{4+kx_{0}^2}}}\nonumber\\(e^{\frac{4ax_{0}}{\sqrt{4+kx_{0}^2}}}(4ax_{0}^2-M^2(4+kx_{0}^2)+Q^2 (4+kx_{0}^2))\nonumber\\-4a^2 x_{0}^2 \Lambda)((4+kx_{0}^2)(e^{\frac{4ax_{0}}{\sqrt{4+kx_{0}^2}}}(4aM^2x_{0} {\sqrt{4+kx_{0}^2}}\nonumber\\-4a Q^2x_{0} {\sqrt{4+kx_{0}^2}} +M^3 (4+kx_{0}^2)\nonumber\\-M(-4a^2x_{0}^2+Q^2(4+kx_{0}^2)))-4a^2Mx_{0}^2 \Lambda)\nonumber\\+16a^3x_{0}^3(2ax_{0}+M {\sqrt{4+kx_{0}^2}}) \Lambda)],
\end{eqnarray}
It is found that
\begin{eqnarray}
\Gamma^{x(-)}_{vv}=[e^{\frac{4ax_{0}}{\sqrt{4+kx_{0}^2}}}(4a^2x_{0}^2+4aMx_{0} {\sqrt{4+kx_{0}^2}}+M^2\nonumber\\ (4+kx_{0}^2)-Q^2 (4+kx_{0}^2))+4a^2x_{0}^2 \Lambda]^{-5}\nonumber\\\times[16a^3 e^{\frac{12ax_{0}}{\sqrt{4+kx_{0}^2}}}x_{0}^{4}{\sqrt{4+kx_{0}^2}}(e^{\frac{4ax_{0}}{\sqrt{4+kx_{0}^2}}}(-4a^2 x_{0}^2\nonumber\\+M^2 (4+kx_{0}^2)-Q^2 (4+kx_{0}^2))+4a^2x_{0}^2 \Lambda)\nonumber\\((4+kx_{0}^2)(e^{\frac{4ax_{0}}{\sqrt{4+kx_{0}^2}}}(4aM^2 x_{0}\sqrt{4+kx_{0}^2}\nonumber\\-4aQ^2 x_{0} \sqrt{4+kx_{0}^2}+M^3 ({4+kx_{0}^2})\nonumber\\-M(-4a^2 x_{0}^2 +Q^2(4+kx_{0}^2)))-4a^2Mx_{0}^2 \Lambda)\nonumber\\+16a^3 x_{0}^3(2ax_{0}+M \sqrt{4+kx_{0}^2})\Lambda)],~~
\end{eqnarray}

Thus,
\begin{eqnarray}
K^{-}_{\tau \tau}=-[16a^2 e^{\frac{8ax_{0}}{\sqrt{4+kx_{0}^2}}} x_{0}^2 ((4+kx_{0}^2)(e^{\frac{4ax_{0}}{\sqrt{4+kx_{0}^2}}}\nonumber\\(4a M^2 x_{0} \sqrt{4+kx_{0}^2}-4aQ^2 x_{0} \sqrt{4+kx_{0}^2}\nonumber\\+M^3 (4+kx_{0}^2)-M(-4a^2x_{0}^2+Q^2(4+kx_{0}^2)))\nonumber\\-4a^2 Mx_{0}^2 \Lambda)+16a^3 x_{0}^3(2ax_{0}+M \sqrt{4+kx_{0}^2})\nonumber\\ \Lambda)]\times[\sqrt{4+kx_{0}^2}(e^{\frac{4ax_{0}}{\sqrt{4+kx_{0}^2}}}(-4a^2x_{0}^2 \nonumber\\+M^2(4+kx_{0}^2)-Q^2 (4+kx_{0}^2))\nonumber\\+4a^2x_{0}^2 \Lambda)(e^{\frac{4ax_{0}}{\sqrt{4+kx_{0}^2}}}(4a^2x_{0}^2 +4aMx_{0} \sqrt{4+kx_{0}^2} \nonumber\\+M^2 (4+kx_{0}^2)-Q^2(4+kx_{0}^2))+4a^2x_{0}^2 \Lambda)^2]^{-1},
\end{eqnarray}
Similarly the extrinsic curvature arising from the exterior $FRW$ region is calculated and we find,
\begin{eqnarray}
K^{+}_{\tau \tau}=0,
\end{eqnarray}

In order to match $K^{-}_{\tau \tau}$ and $K^{+}_{\tau \tau}$, this would simply imply,
\begin{eqnarray}
(4+kx_{0}^2)(e^{\frac{4ax_{0}}{\sqrt{4+kx_{0}^2}}}(4a M^2 x_{0} \sqrt{4+kx_{0}^2}-4aQ^2 x_{0}\nonumber\\ \sqrt{4+kx_{0}^2}+M^3 (4+kx_{0}^2)-M(-4a^2x_{0}^2\nonumber\\+Q^2(4+kx_{0}^2)))-4a^2 Mx_{0}^2 \Lambda)+16a^3 x_{0}^3\nonumber\\(2ax_{0}+M \sqrt{4+kx_{0}^2})\Lambda=0,
\end{eqnarray}

Hence the metric as well as the extrinsic curvature are continuous at the boundary surface.

\section{Further discussion on surface continuity:}\label{sec7}

We now prove the surface continuity alternatively. Let the surface be discontinuous. Then on
the contrary, the discontinuity in the extrinsic curvature determine the surface stress energy and surface tension of the junction surface at $x = x_{0}$ where the surface stress-energy tensor components are determined [\cite{a},\cite{b},\cite{d},\cite{e}].

Let,
\begin{eqnarray}
f(x)=\frac{(1-\frac{M^2
		(1+\frac{kx^2}{4})}{a^{2}x^{2}}+\frac{Q^{2}(1+\frac{kx^2}{4})}{a^{2}x^2}
	-\Lambda e^{-\frac{2ax}{\sqrt{1+\frac{kx^2}{4}}}})^2}
{[(1+\frac{M {\sqrt{1+\frac{kx^2}{4}}}}{ax})^2-\frac{Q^2
		(1+\frac{kx^2}{4})}{a^{2}x^2}+\Lambda
	e^{-\frac{2ax}{\sqrt{1+\frac{kx^2}{4}}}}]^2},\nonumber\\
 \label{eqn.71}
\end{eqnarray}
\begin{eqnarray}
g(x)=\frac{a^2}{(1+\frac{kx^2}{4})^2}[(1+\frac{M
	{\sqrt{1+\frac{kx^2}{4}}}}{ax})^2 \nonumber\\
-\frac{Q^2 (1+\frac{kx^2}{4})}{a^{2}x^2}+\Lambda
e^{-\frac{2ax}{\sqrt{1+\frac{kx^2}{4}}}}]^2,
\end{eqnarray}
\begin{eqnarray}
h(x)=\frac{a^2 x^2}{(1+\frac{kx^2}{4})^2}[(1+\frac{M
	{\sqrt{1+\frac{kx^2}{4}}}}{ax})^2 \nonumber\\
-\frac{Q^2 (1+\frac{kx^2}{4})}{a^{2}x^2}+\Lambda
e^{-\frac{2ax}{\sqrt{1+\frac{kx^2}{4}}}}]^2,
\end{eqnarray}
Hence,
\begin{eqnarray}
f'(x)=[(4+kx^2)^{-3/2}(e^{\frac{4ax}{\sqrt{4+kx^2}}}(4a^2 x^2\nonumber\\+4a Mx \sqrt{4+kx^2}+M^2 (4+kx^2)-Q^2\nonumber\\ (4+kx^2))+4a^2 x^2 \Lambda)^{-3}][32 ae^{\frac{4ax}{\sqrt{4+kx^2}}}\nonumber\\(e^{\frac{4ax}{\sqrt{4+kx^2}}}(4ax^2-M^2(4+kx^2)+Q^2 (4+kx^2))\nonumber\\-4a^2 x^2 \Lambda)((4+kx^2)(e^{\frac{4ax}{\sqrt{4+kx^2}}}(4aM^2x {\sqrt{4+kx^2}}\nonumber\\-4a Q^2x {\sqrt{4+kx^2}} +M^3 (4+kx^2)\nonumber\\-M(-4a^2x^2+Q^2(4+kx^2)))-4a^2Mx^2 \Lambda)\nonumber\\+16a^3x^3(2ax+M {\sqrt{4+kx^2}}) \Lambda)],
\end{eqnarray}

The jump of the extrinsic curvature components at the surface $x = x_{0}$, is associated
with the surface energy density \cite{c} as,
\begin{eqnarray}
\lambda(x_{0})=-\frac{1}{8 \pi} {\frac{h'(x_{0})}{h(x_{0})}} {\sqrt{\frac{1+\dot{x_{0}}^2 g(x_{0})}{g(x_{0})}}}
\end{eqnarray}
and the surface pressure as,
\begin{eqnarray}
P(x_{0})=\frac{1}{8 \pi} {\sqrt{\frac{1+\dot{x_{0}}^2 g(x_{0})}{g(x_{0})}}} [2 \ddot{x_{0}}+{\dot{x_{0}}}^2 (\frac{f'(x_{0})}{f(x_{0})}+\frac{g'(x_{0})}{g(x_{0})}) \nonumber\\+ \frac {f'(x_{0})}{f(x_{0})g(x_{0})}],~~~~~
\end{eqnarray}

For a static configuration of radius $x_{0}$, we obtain (assuming $\dot{x}_{0} = 0$ and $\ddot{x}_{0} = 0$)

\begin{eqnarray}
\lambda(x_{0})=-\frac{1}{8 \pi} {\frac{h'(x_{0})}{h(x_{0})}} {\sqrt{\frac{1}{g(x_{0})}}}\nonumber\\
= [4 \pi x_{0} (4+kx_{0}^2)^{\frac{3}{2}}(e^{\frac{4ax_{0}}{\sqrt{4+kx_{0}^2}}}(4a^2x_{0}^2 +4aMx_{0} \nonumber\\\sqrt{4+kx_{0}^2} +M^2 (4+kx_{0}^2)-Q^2(4+kx_{0}^2))\nonumber\\+4a^2x_{0}^2 \Lambda)]^{-1}[4ae^{\frac{4ax_{0}}{\sqrt{4+kx_{0}^2}}}kM x_{0}^3 (4+kx_{0}^2)\nonumber\\+e^{\frac{4ax_{0}}{\sqrt{4+kx_{0}^2}}} (M^2-Q^2)(4+kx_{0}^2)^{\frac{5}{2}}+4a^2x_{0}^2(-4+kx_{0}^2)\nonumber\\{\sqrt{4+kx_{0}^2}}(e^{\frac{4ax_{0}}{\sqrt{4+kx_{0}^2}}}+\Lambda)+64a^3 x_{0}^3 \Lambda]\nonumber\\\times[\frac{(4+kx_{0}^2)}{4a (1+\frac{M \sqrt{4+kx_{0}^2} }{ax_{0}}+\frac{(M^2-Q^2)(4+kx_{0}^2)}{4a^2 x_{0}^2}+\Lambda e^{-\frac{4ax_{0}}{\sqrt{4+kx_{0}^2}}})}],~~~~~
\end{eqnarray}
and
\begin{eqnarray}
P(x_{0})=\frac{1}{8 \pi} {\sqrt{\frac{1}{g(x_{0})}}} [\frac {f'(x_{0})}{f(x_{0})g(x_{0})}]\nonumber\\
=-[\pi (e^{\frac{4ax_{0}}{\sqrt{4+kx_{0}^2}}}(-4a^2 x_{0}^2 +M^2 (4+kx_{0}^2)-Q^2(4+kx_{0}^2))\nonumber\\+4a^2x_{0}^2 \Lambda)(e^{\frac{4ax_{0}}{\sqrt{4+kx_{0}^2}}}(4a^2x_{0}^2\nonumber+4aMx_{0} \sqrt{4+kx_{0}^2}\nonumber\\ +M^2 (4+kx_{0}^2)-Q^2(4+kx_{0}^2))+4a^2x_{0}^2 \Lambda)^3]^{-1}\nonumber\\\times[a^2 e^{\frac{12ax_{0}}{\sqrt{4+kx_{0}^2}}} x_{0}^4 (4+kx_{0}^2)^{\frac{3}{2}}(1+\frac{M \sqrt{4+kx_{0}^2} }{ax_{0}}\nonumber\\+\frac{(M^2-Q^2)(4+kx_{0}^2)}{4a^2 x_{0}^2}+\Lambda e^{-\frac{4ax_{0}}{\sqrt{4+kx_{0}^2}}})^{-1} ((4+kx_{0}^2)\nonumber\\(e^{\frac{4ax_{0}}{\sqrt{4+kx_{0}^2}}}(4a M^2 x_{0} \sqrt{4+kx_{0}^2}-4aQ^2 x_{0}\sqrt{4+kx_{0}^2}+M^3\nonumber\\ (4+kx_{0}^2)-M(-4a^2x_{0}^2+Q^2(4+kx_{0}^2)))-4a^2 Mx_{0}^2 \Lambda)\nonumber\\
+16a^3 x_{0}^3(2ax_{0}+M \sqrt{4+kx_{0}^2})\Lambda)],~~~~~
\end{eqnarray}
which significantly reduces to $P(x_{0})=0$ using eqn.(84). The vanishing surface pressure thus proves the metric continuity at the boundary surface, i.e on the horizon $x=x_{0}$, as envisaged.

\section{Conclusions:}\label{conclusions}

We thus study a charged, non-rotating, spherically symmetric black hole which has cosmological constant $\Lambda$ (Reissner-Nordstr\"{o}m+$\Lambda$),  active
gravitational  mass $M$ and electric charge $Q$ in
exterior Friedman-Robertson-Walker (FRW) universe. The
Einstein-Maxwell equations of the RN+$\Lambda$ black hole embedded
in the FRW background are solved. As a procedure, we have started with a (2+1)-d RN+$\Lambda$ black hole and then performed
a simple transformation only under suitable conditions to obtain a metric which matches with the
exterior Friedman-Robertson-Walker universe universe and
also derived a negative cosmological constant inside the black hole. New
classes of exact solutions of the charged
black hole are found. Literature reveals that there are
three possible black hole solutions where the cosmological constant is
(1) positive (2) negative and (3) zero. The cosmological constant found
negative inside the black hole is also confirmed by the geodesic equations. Here,
the cosmological constant is dependent on $R, Q$ and
$a(v)$ which correspond to the areal radius, charge, of the black hole and the
scale factor of the universe respectively. The century-old problem of describing a gravitationally
bound system in an expanding universe in the frame-set of general relativity has seen many attempts to find a solution.
Assuming that scale factor does not alter with the metric transformation, we find a maximum limit of the
universal expansion. Despite its apparent simplicity, a full understanding of the mechanisms involved when general and realistic systems
are considered has yet to be found. We also observe that the size, mass and charge
of the black hole is affected by the expansion of the universe.
An important observation is that, for an observer at infinity, both the mass
and charge of black hole increase with the contraction of the universe and
decrease with the expansion of the universe. The cosmological constant
has been found to be negative in a previous work too \cite{landry12}. The
AdS/CFT correspondence tells us that the case $\Lambda<0$  is still worthy of
consideration. In future we plan to study the stability of such black hole with
cosmological constant in an expanding universe.

We justify the use of two different methods for matching spacetimes. Boundary and matching conditions with the exterior FRW universe are studied in section IV., to arrive at a conclusion that the cosmological constant can have a negative value inside the black hole. However, the Darmois-Israel matching conditions have been studied in section VI., to deduce that the metric as well as the extrinsic curvature are continuous at the boundary surface. Alternatively the vanishing surface pressure also proves the metric continuity at the boundary surface, i.e on the horizon $(x = x_0)$.

\section*{ACKNOWLEDGEMENTS}

SI is thankful to Harish-Chandra Research Institute,
Allahabad for providing Post Doctoral research support. SI is thankful to the authority of Inter-University
Centre for Astronomy and Astrophysics, Pune, India for
providing Visiting support. FR is thankful to the authority of Inter-University
Centre for Astronomy and Astrophysics, Pune, India for
providing Visiting Associateship.  This work  is a part of the project submitted by FR in SERB, DST.

\end{document}